# Designing for Collaborative Sensemaking: Using Expert & Non-Expert Crowd

**Nitesh Goyal**

Information Science, Cornell University
ngoyal@cs.cornell.edu

## Abstract

Crime solving is a domain where solution discovery is often serendipitous. Unstructured mechanisms, like Reddit, for crime solving through crowds have failed so far. Mechanisms, collaborations, workflows, and micro-tasks necessary for successful crime solving might also vary across different crimes. Cognitively, while experts might have deeper domain knowledge, they might also fall prey to biased analysis. Non-experts, while lacking formal training, might instead offer non-conventional perspectives requiring direction. The analytical process is itself an iterative process of foraging and sensemaking. Users would explore to broaden solution space and narrow down to a solution iteratively until identifying the global maxima instead of local maxima. In this proposal, my research aims to design systems for enabling complex sensemaking tasks that require collaboration between remotely located non-expert crowds with expert crowds to compensate for their cognitive challenges and lack of training. This would require better understanding of the structure, workflow, and micro-tasks necessary for successful collaborations. This proposal builds upon previous work on collaborative sensemaking between remote partners in lab experiments and endeavors to scale it across multiple team members, with varying expertise levels.

## Motivation

While sensemaking has been studied in the past, designing interfaces for relatively complex sensemaking where experts and non-experts may collaborate remains a challenge. Further, if we could leverage human cognition for collaborative sensemaking, crowds may help us better solve unstructured problems where traditional computational techniques have failed. In particular, researching how to design for experts and non-experts in the crime-solving domain, where solutions are often found through serendipity instead of rules, might offer us insights into how to best utilize crowd expertise, and leverage it to solve otherwise hard problems.

## Background and Related Work

Crowdsourcing for somewhat complex tasks has been pursued in the past. Collaborative document editing in Soylent (Bernstein et al. 2010), creating taxonomy of colors in Cascade (Chilton et al. 2013), suggesting a travel itinerary under constraints using Mobi (Zhang et al, 2012), and mining sentiments by crowds for text analytics in OpinionBlocks (Hu et al, 2013) are some recent forays where crowdsourcing has shown to be performant and/or efficient. However, more open-ended domains like crime solving, requiring serendipitous discovery of clues and criminals, have yet to be crowd-sourced successfully.

As number of workers and associated workflows grow in complexity, crowdsourcing can be challenging (Bernstein 2010). Crowdsourcing for complex workflows has been pursued also. For example, CrowdForge explains how map-reduce framework popularized by Google may be used to partition bigger complex tasks into smaller tasks dynamically by workers (Kittur et al,. 2005). Further, Malone et al 's aggregation dimension suggests that crowdworkers can either work alone independently or depend upon each other to work together (Malone et al, 2005)]. TurKit can further help decide what to present to each worker such that the flow of results of tasks between dependent workers can be controlled (Little et al 2010).

As such crowd-workflows become complex, researchers must identify the level of crowd-supervision needed for optimal output. Turkomatic was designed based on price-divide-loop such that real time visualization of the workflow-design is evident because unsupervised crowds failed to produce proper workflows resulting in a less than optimal output (Kulkarni et al 2012). On the other hand, supervised crowds in a conversational-agent, Chorus (Lasecki et al. 2013), made users believe that a single user exist behind Chorus. Instead, Chorus employs multiple crowd workers who collectively create response possibilities, such that Crowd workers can learn and remember collectively.

Alternatively, Kulkarni shows a computational method of identifying Experts in a crowd who subsequently utilize non-Experts to perform the tasks, in Wish (Kulkarni, et al 2014). Other researchers have pursued task-routing based on expertise level (Bragg et al. 2014). To summarize, while relatively complex tasks and workflows using crowds have been attempted, we have yet been unable to design a system that may structure non-experts (lesser trained crowds) and experts (trained workers) together in an interface to solve complex challenges like crime solving.

## Proposed Research

The core aim of this research is to pursue a user centered design approach to designing a web-interface and an underlying system that may enable collaborations between experts and non-experts, and within non-experts. I hypothesize that such an interface for solving carefully broken down micro-tasks would help leverage distributed human cognition to solve complex tasks like crimes. To pursue this research, at least the following two important research questions need to be pursued:

### Research Questions
**First, understanding how to break up the task and data (materials) is important to imitate a realistic crime-solving scenario.** This a wide landscape, where in one direction, one may consider sharing the entire dataset with the crowds and then enable the crowds to deduce the partition mechanism for the dataset and create associated workflows. However, this requires expertise. On the other hand, an expert might already partition the dataset into smaller datasets that are visible uniquely to each crowd worker. However, this prevents crowds from directly finding global maxima, since the broader overview is not visible to the workers. Perhaps, the solution is somewhere in between the two extremes and this knowledge will help find answer to the next question.

**Second, identifying the workflow for user collaborations to enable the best solution discovery is important to overcome biased knowledge generation.** Should the users perform solo work initially, and then collaborate on the fruits of solo labor? Or vice versa? Or, a combination hitherto. It is unclear how to enable collaboration for a fuller attention spread across all the possible solutions? Does this process remain static, or should the system computationally identify opportune moments to suggest collaborations based on the user performance? It is yet unclear, the role of algorithmic aids, and collaboration with crowds to best promote unbiased solution discovery. Further, online collaborations between crowd members require operational stability in case of drop-offs, or inactivity.

### Planned Methodology
I plan to integrate my findings based on a mixed-methods study. First, I will understand how trained-non-experts (trained students, through video and usage-log analysis) solve complex problems singularly and collaboratively. Consequently, I will extract important features that result in success and failure in problem solving. Based on these features, I propose to create a web-interface for collaborative problem solving. Finally, I will design a study to validate whether the identified features (reflected as expertise) lead to success or failure with non-expert crowds, and a mix of expert + non-expert crowds? So, based on the iterative nature of design process, my proposed solution would involve multiple iterations and steps before I design the final interface:
Step 1. Understand role of currently used features for solo sensemaking.
Step 2. Explore effects of information-sharing collaborative sensemaking.
Step 3. Extract features to identify micro-tasks, and workflows for success.
Step 4. Design web-interface for experts and non-experts to collaborate.
Step 5. Design a set of user-studies to measure user-experience, and performance achieved by non-experts with the web-interface at solving crimes.

### Progress so far
I have completed three iterations of system building of SAVANT tool to support solo (Goyal et al. 2012) and collaborative sensemaking (Goyal et al. 2013)(Goyal et al. 2014)(Goyal et al. 2016) to better understand role of different design features:

In Iteration 1 (Step 1), I tested the utility of system-generated visualization of data links and a notepad for collecting annotations, and found system-generated visualizations to be significantly important in solving crimes (Goyal et al. 2012).

In Iteration 2 (Step 2), I explored value of implicitly sharing insights by self-created visualizations of annotations, without explicitly pushed/requested information by collaborators. When implicit sharing of notes and self-created visualization of these notes was available, users identified more clues (Goyal et al. 2013)(Goyal et al. 2014).

In Iteration 3 (Step 2), I explored value of visualizing real-time sensemaking translucence to reduce biased analysis using NLP on implicitly shared notes, and explicit chat channel. With sensemaking translucence, users improved task-performance from previous work (Goyal et al. 2013)(Goyal et al. 2014) by identifying the serial killer significantly more(Goyal et al. 2016).

In Step 3 I finished conducting a qualitative video-analysis, and usage-log analysis of the actions performed by successful and unsuccessful pairs in Step 2. Based on video-analysis, 3 design goals seem promising for success: *externalizing insights; shoe-boxing visually; and iterating over previously collected information*.

For Step 3, I am also identifying user-actions, based on interface-log analysis, when pursued multiple times by users would lead to successful resolution of the task.

**Next Steps**

Based on preliminary findings, I am designing the web-interface (SAVANT) for non-experts and experts with recommended steps associated with success. Based on these findings, I'd be better equipped with knowledge of micro-tasks that would enable success in task-resolution. So, I'd propose using the lessons learnt to design the next SAVANT version where users using the full SAVANT suite might be able to collaborate and auto-direct micro-tasks to crowds that would support/challenge their own insights and help resolve the crime-solving task.

However, what remains to be pursued, is thematically segregated as follows:
1. Identify the best workflows and aggregation mechanisms that support collaboration between expert and non-expert crowd workers to identify the serial killer.
2. Identify system-generated micro-tasks vs. manually generated micro-tasks and associated cost vs. benefit ratio.
3. Understand how to generalize the findings from this work.

## Challenges

The following challenges need to be overcome for the success of this work. So, feedback in the following directions and potential experiment designs would be useful:
1. *Identifying optimal data/task set-up that balances utility of the task with resemblance to the real-world situations*: I am hoping that deliberation on how to better characterize the task for crowd would be beneficial.
2. *Identifying an aggregation mechanism that enables individual crowd workers to not just identify a killer, but the serial killer across multiple crime cases*: Aggregated results may be reiterated through the crowds.
3. *Identifying workflows that enable collaboration between experts and non-experts for optimal knowledge generation and dissemination*: Procedural and temporal collaboration effects need to be studied.
4. *Identifying the balance between computer and human computation*: Understanding, when and how to aid the crowds will help improve task performance, and perhaps user satisfaction.

# Appendix

I want to participate in the Doctoral consortium at this point because after multiple controlled studies in lab where careful modifications have yielded directions on how to improve collaborative sensemaking between pairs, I would like to scale my work to larger worker pool. To this end, I have identified important research questions that would help scaling my prior research to wider user population and support generalization of my work. I am hoping to get feedback on how to pursue these questions because I have already proposed this work. I am in the midst of creating a lightweight front-end of crowd worker's web interface as an extension of prior work with SAVANT. The back-end and logic design of this system will be generated, over remaining Fall 2015 and early Spring 2016, based on the feedback received at HCOMP 2015. Feedback will also help design subsequent experiments to be run in late Spring 2016. I am hoping to write the results as my dissertation over Summer 2016 with an expected defense date in mid September 2016.